\documentclass[fleqn,usenatbib]{mnras}

\usepackage{newtxtext,newtxmath}
\usepackage[T1]{fontenc}
\usepackage{ae,aecompl}
\usepackage{url}
\usepackage{graphicx}
\usepackage{xspace}
\usepackage{amsfonts}
\usepackage{pifont}
\usepackage{textcomp}
\usepackage{float}
\usepackage{subfig}
\usepackage{multirow}
\usepackage{adjustbox}

\title[TESS observations of MQ Dra]{Detection of an energetic flare from the M5V secondary star in the Polar MQ Dra}

\author[G. Ramsay et al.]{
Gavin Ramsay$^{1}$\thanks{E-mail: gavin.ramsay@armagh.ac.uk}, Pasi Hakala$^{2}$, Matt A. Wood$^{3}$
\\
$^{1}$Armagh Observatory and Planetarium, College Hill, Armagh, BT61 9DG, UK\\
$^{2}$Finnish Centre for Astronomy with ESO (FINCA), Quantum, University of Turku, FI-20014, Finland\\
$^{3}$Department of Physics and Astronomy, Texas A\&M University-Commerce, 
Commerce, TX 75429-3011, USA\\
}

\date{Accepted 2021 April 19. Received 2021 April 19; in original form 2021 March 02}

\begin{document}
\label{firstpage}
\pagerange{\pageref{firstpage}--\pageref{lastpage}}

\outer\def\gtae {$\buildrel {\lower3pt\hbox{$>$}} \over 
{\lower2pt\hbox{$\sim$}} $}
\outer\def\ltae {$\buildrel {\lower3pt\hbox{$<$}} \over 
{\lower2pt\hbox{$\sim$}} $}
\newcommand{\Msun}{$M_{\odot}$}
\newcommand{\lsun}{$L_{\odot}$}
\newcommand{\Rsun}{$R_{\odot}$}
\newcommand{\solar}{${\odot}$}
\newcommand{\kep}{\sl Kepler}
\newcommand{\ktwo}{\sl K2}
\newcommand{\tess}{\sl TESS}
\newcommand{\swift}{\it Swift}
\newcommand{\Porb}{P_{\rm orb}}
\newcommand{\nuorb}{\nu_{\rm orb}}
\newcommand{\eplus}{\epsilon_+}
\newcommand{\eminus}{\epsilon_-}
\newcommand{\cd}{{\rm\ c\ d^{-1}}}
\newcommand{\MdotL}{\dot M_{\rm L1}}
\newcommand{\Mdot}{$\dot M$}
\newcommand{\Mdsolar}{\dot{M_{\odot}} yr$^{-1}$}
\newcommand{\Ldisk}{L_{\rm disk}}
\newcommand{\src}{KIC 9202990}
\newcommand{\ergscm} {erg s$^{-1}$ cm$^{-2}$}
\newcommand{\rchi}{$\chi^{2}_{\nu}$}
\newcommand{\chisq}{$\chi^{2}$}
\newcommand{\pcmsq} {cm$^{-2}$}

\maketitle

\begin{abstract}
MQ Dra is a strongly magnetic Cataclysmic Variable whose white dwarf
accretes material from its secondary star through a stellar wind at a
low rate. {\tess} observations were made of MQ Dra in four sectors in
Cycle 2 and show a short duration, high energy flare ($\sim10^{35}$
erg) which has a profile characteristic of a flare from the M5V
secondary star.  This is one of the few occasions where an energetic
flare has been seen from a Polar. We find no evidence that the
  flare caused a change in the light curve following the event and
  consider whether a coronal mass ejection was associated with the
  flare. We compare the frequency of energetic flares from the
  secondary star in MQ Dra with M dwarf stars and discuss the overall
  flare rate of stars with rotation periods shorter than 0.2 d and how
  such fast rotators can generate magnetic fields with low
  differential rotation rates.
\end{abstract}

\begin{keywords}
Physical data and processes: accretion, accretion discs -- stars: binaries close -- 
stars: activity -- stars: flare -- stars: low-mass -- stars: magnetic fields 
\end{keywords}

\section{Introduction}

Polars, or AM Her stars, are compact binaries where a strongly
magnetic ($B>10$ MG) white dwarf accretes material through Roche Lobe
overflow from a dwarf secondary star. They are members of the wider
group of compact accreting binaries called cataclysmic variables
(CVs). Unlike many CVs which show dwarf novae outbursts, the strong
magnetic field of the white dwarf prevents the formation of an
accretion disc around the white dwarf and material gets channelled
onto one or both of its magnetic poles. Strong shocks get formed above
the photosphere of the white dwarf which results in X-ray emission and
cyclotron emission which is strongly beamed \citep[see][for an early
  review]{Cropper1990}.

Polars can exhibit different accretion states where they appear in
'high', 'intermediate' or 'low' states. Some systems, such as EF Eri,
have spent much of the last two decades in a low state apart from
short duration high states \citep[e.g.,][]{Schwope2007}. Although
star-spots near the Roche Lobe on the secondary star may suppress mass
transfer \citep{LivioPringle1994}, it is still unclear why these state
transitions occur. However, when a system is in a low or intermediate
state, we have the opportunity to observe activity on the secondary
star. In addition there is a small sample of Polars which accrete at a
very low rate and whose secondary is not thought to be filling their
Roche Lobe, so that accretion takes place through a stellar wind
\citep[e.g.,][]{Schwope2002}. As a result they are excellent targets to study
activity of the secondary star.

The mass and spectral type of the secondary star in a CV is related to
the binary orbital period, with longer period systems (\gtae5 hrs)
having stars with late K spectral type, whilst those with periods
shorter than $\sim$2 hrs having stars with spectral type M5V or later
\citep[e.g.,][]{Knigge2006}. The polars therefore have secondary stars
which straddle the boundary point (M4V) where stars are fully
convective (later) or partially convective/radiative (earlier). Since
fully convective stars also show evidence of stellar activity, it was
thought that they must generate and maintain their magnetic field in a
different way from Solar type stars. However, more recently,
  observations \citep[e.g.][]{WrightDrake2016} show that slowly
  rotating fully convective M dwarfs have X-ray properties identical
  to slowly rotating Solar type stars, whilst theoretical work
  \citep[e.g.][]{Yadav2016} has indicated that the nature of the
  dynamo mechanism changes as fully convective stars age and slow
  down. Observations made using {\kep}, {\ktwo}
  \citep[e.g.][]{Raetz2020} and {\tess} \citep[e.g.][]{Gunther2020}
  enables the flaring rates of many low mass stars to be
  determined which helps address the question of how stable magnetic
  fields are generated in stars in general.

One low accretion rate Polar whose M5V secondary appears to underfill
its Roche Lobe and accrete via a wind is MQ Dra (SDSS J1553+55) which
shows very prominent broad emission features which were attributed to
cyclotron radiation, which allowed the magnetic field strength to be
determined \citep[$\sim$60 MG,][]{Szkody2003,Schmidt2005}. Subsequent
modelling indicated the mass transfer rate was very low
($\dot{M}\sim6\times10^{-14}$ {${M_{\odot}}$ yr$^{-1}$}), implying a
very low specific accretion rate, ($\dot{m}\sim0.001-0.01$ g cm$^{-2}$
s$^{-1}$), and hence no strong shock would form. In contrast, the low
accretion rate and the high magnetic field gives rise to widely
separated and prominent cyclotron harmonics in the optical
\citep{Szkody2004,Schmidt2005}.  The parallax from the {\sl Gaia} EDR3
catalogue \citep{Gaia2020} is 5.51$\pm$0.06 mas, implying a distance
of 181.7$^{+3.5}_{-3.3}$ pc.

Optical photometry of MQ Dra has shown that the amplitude and shape of
the data folded on the orbital period is highly dependent on the
filter used in the observations. The emission observed over the
UV-optical-IR bands is a sum of beaming of the cyclotron emission; the
white dwarf and the ellipsoidal modulation due to the changing viewing
angle of the secondary star. In the $R$ band the amplitude reaches a
peak-to-peak variation of $\sim$1 mag and only one peak over the
orbital period \citep{Szkody2008}. However, a double peak is seen in
the $B$ and $I$ bands which could be a combination of cyclotron
beaming and the ellipsoidal modulation of the secondary.  In this
short paper we present the photometric data of MQ Dra obtained using
the {\tess} mission during Cycle 2 when it observed the northern
ecliptic hemisphere. In particular, we identify an energetic stellar
flare and discuss the activity levels on secondary stars in polars and
compare them with field M dwarfs.

\section{{\tess} observations}
\label{tess}

{\tess} was launched in April 2018 and consists of four 10.5 cm
telescopes that observe a 24$^{\circ}\times96^{\circ}$ strip (known as
a {\it sector}) of sky for $\sim$28~d \citep[see][for
  details]{Ricker2015}.  Between July 2018 and June 2019, {\tess}
covered most of the southern ecliptic hemisphere (Cycle 1) and between
July 2019 and June 2020 covered most of the northern ecliptic
hemisphere (Cycle 2). Although there is a band along the ecliptic
plane which was not observed, at the ecliptic poles there is a
continuous viewing zone where stars can be observed for $\sim$1
yr. Each `full-frame image' has an exposure time of 30 min. However,
in each sector, photometry with a cadence of 2 min is obtained, with
most targets being selected from the community via a call for
proposals.

\begin{figure*}
  \begin{center}
  \includegraphics[width=0.95\textwidth]{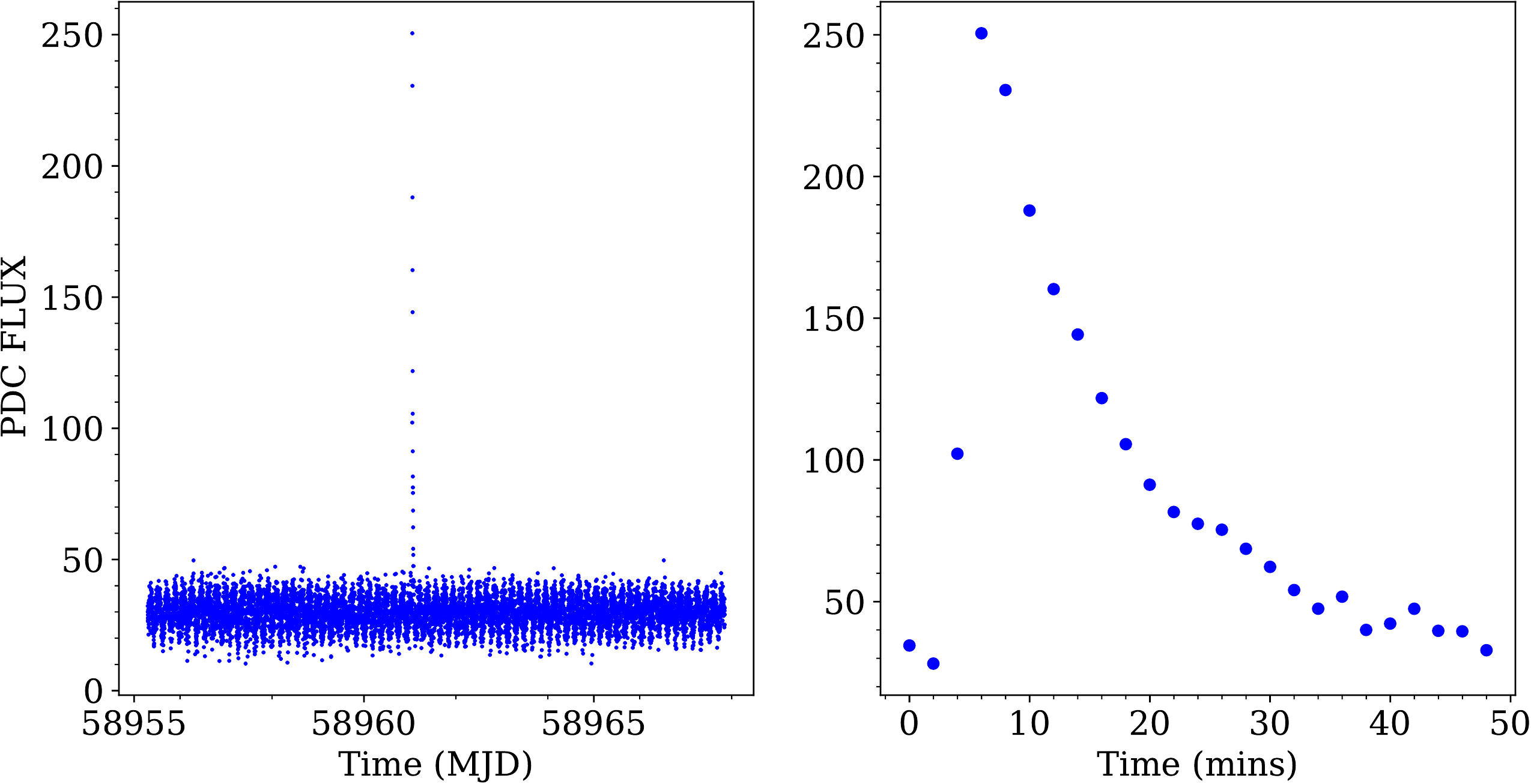}    
\vspace{2mm}
  \caption{The left hand panel shows the light curve taken from the
    first half of {\sl TESS} sector 24 and the right hand panel shows the
    light curve focused on the flare where we have converted the time
    to show the time in mins from close to the start of the flare.}
    \label{figure1}
    \end{center}
\end{figure*}

MQ Dra (TIC 161744828) was observed in Cycle 2 during sectors 16, 22,
23 and 24 in 2 min cadence (see Table \ref{table1}).  We downloaded
the calibrated lightcurves of our targets from the MAST data
archive\footnote{\url{https://archive.stsci.edu/tess/}}. We used the
data values for {\tt PDCSAP\_FLUX}, which are the Simple Aperture
Photometry values, {\tt SAP\_FLUX}, after correction for systematic
trends. We removed photometric points which did not have {\tt
  QUALITY=0} flag. There is a prominent modulation on the known
orbital period of 0.182985 d (4.39 hr) \citep{Szkody2008} which is
apparent in the data from all sectors. The relative amplitude appears
to decline over the four sectors (0.63 mag sector 16, 0.49 mag sector
22, 0.38 mag sector 23 and 0.34 mag sector 24). It is likely that this
is an instrumental affect (there are three separate Camera and CCD
combinations for the observations for MQ Dra).

Examining the light curves from each sector, the observation from
sector 24 showed a clear and short duration increase in brightness,
from a mean level of 30 ct/s to a peak of 250 ct/s (left hand Figure
\ref{figure1}). We show the detail of the event in the right hand of
Figure \ref{figure1}: it shows a profile which has a rapid rise to
maximum flux followed by a more gradual decline, the whole event
lasting $\sim$40 min. The profile is characteristic of a stellar flare
which has an amplitude of 2.3 mag in the {\tess} pass-band (600 -- 1000 nm). 
There is no indication that the
orbital brightness profile changed over the following orbital cycles.

Given that the pixel size of the {\tess} cameras is 21 arcsec per
  pixel, it is possible that the flare we detect originated from a
  spatially nearby star. Indeed, \citet{Jackman2021} made a systematic
  study of flares from M dwarfs using {\kep} and {\tess} data and
  conclude that for {\tess} data there is a 5.8 percent chance of a
  false positive flare event from the object of interest due to
  spatially nearby stars. We used {\tt tpfplotter} \citep{Aller2020}
  to overlay the position of stars in the Gaia DR2 catalogue onto an
  image derived from a TESS Target Pixel File. This indicates which,
  if any, stars were in the aperture mask used to extract the light
  curve of the target, which is 3 pixels in size for MQ Dra. There are
  no stars down to $G$=21 mag within the aperture mask of MQ
  Dra. Further we found no evidence that the center of the stellar
  profile of the source observed by {\tess} changed systematically
  over the course of the flare event. We conclude that we identified a
  stellar flare from the secondary star of MQ Dra.

\begin{figure}
  \begin{center}
  \includegraphics[width=0.45\textwidth]{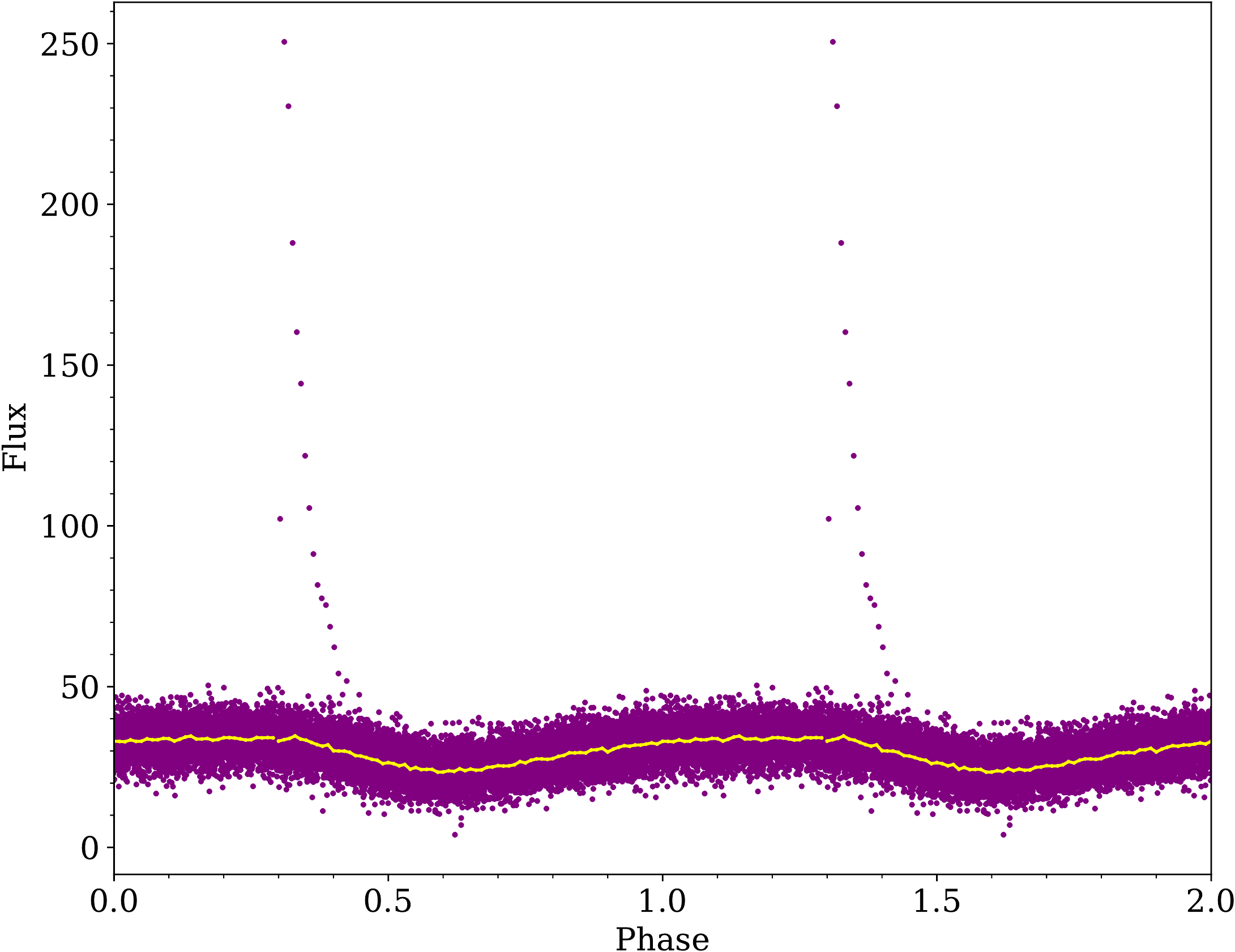}    
\vspace{2mm}
  \caption{The data from sector 24 phased on the definition of
    \citet{Szkody2008} where $\phi$=0.0 marks the inferior conjunction
    of the secondary. This indicates the flare could have illuminated the white dwarf.}
    \label{phase}
    \end{center}
\end{figure}

We can determine the orbital phase by fixing $\phi=0.0$ to match the
flux maximum in the $V$ and $R$ bands from \citet{Szkody2008}, where
this corresponds to the point of inferior conjunction, the point in
the orbit where the secondary is closest to us, i.e., its hemisphere
which is facing away from the white dwarf is facing us. In practise we
align the data so that minimum corresponds to $\phi=0.6$ \citep[see
  Figure 1 of][]{Szkody2008}. In Figure \ref{phase} we show the data
from sector 24 phased on these terms and find the flare occurred over
$\phi\sim0.3$--0.4: this implies the flare could have illuminated the
white dwarf.  There is some uncertainty about the binary inclination,
with \citet{Szkody2008} showing that $i=35^{\circ}$ or $i=60^{\circ}$
give better fits to different aspects of their data. Although there is
no evidence that MQ Dra changed its brightness after the flare it is
possible any mass accretion event would have been seen at UV or X-ray
energies.

\begin{table}
  \begin{tabular}{lcc}
    \hline
    Sector & Start & End \\
    \hline
       16 & 58738.15  2019-09-12T03:37 & 58761.12 2019-10-05T02:50\\
       22 & 58899.43  2020-02-20T10:15 & 58925.99 2020-03-17T23:51\\
       23 & 58929.65  2020-03-21T15:37 & 58954.36 2020-04-15T08:42\\
       24 & 58955.30  2020-04-16T07:08 & 58981.78 2020-05-12T18:45\\
       \hline
  \end{tabular}
  \caption{The start and end times of the {\tess} observations of MQ Dra
    where we show the sector and start and end time in BMJD and
    calendar date.}
  \label{table1}
\end{table}

\section{Flare energy}

The apparent brightness of MQ Dra is a combination of flux from the
white dwarf, the secondary star and cyclotron emission, all of which
vary over the orbital period and can result in an amplitude of 1 mag
in the $R$ band. However, we know that the secondary star has a M5~V
spectral type \citep{Schmidt2005}.

Using Gaia EDR3 \citep{Gaia2020}, we selected those stars within 50 pc
of the Sun and then obtained their bolometric luminosity from the TIC
V8.0 catalogue \citep{Stassun2019} which uses Gaia DR2
\citep{Gaia2018} to determine the luminosities.  Using the work of
Eric
Mamajek\footnote{\url{https://www.pas.rochester.edu/~emamajek/EEM_dwarf_UBVIJHK_colors_Teff.txt}}
\citep{PecautMamjek2013}, we find that a M5 V star has on average a
Gaia colour $(BP-RP)\sim3.35$. Using the Gaia EDR3/TIC sample we find
that the average bolometric luminosity of a M5 V star is
8.8$\times10^{30}$ erg/s (the luminosity of a M4.5 and M5.5 star is
1.3$\times10^{31}$ erg/s and 5.4$\times10^{30}$ erg/s, respectively).

Rather than fit the brightness profile of the flare using a flare
model, which is appropriate for automatically identifying and
  characterising flares from many light curves, we simply summed up
the area under the flare after pre-whitening the light curve to remove
the modulation on a period of 4.39 hr. This results in a flare
luminosity of 2.6$\times10^{34}$ erg.  However, we need to correct
this to take into account that the temperature of the flare can be
considerably hotter than the photosphere of the M dwarf. We assume
that the temperature of the flare is $\sim$12000 K and that the
fraction of the emitted flux which falls within the {\tess} pass-band
is $\sim$0.14 \citep{Schmitt2019}. This implies a correction factor of
$\sim$7 to obtain the bolometric luminosity of the flare implying a
bolometric luminosity of $\sim2\times10^{35}$ erg, with a factor of
2--3 uncertainty to take into account the exact spectral type and
temperature of the flare.

Observations of the Sun show that many flares are associated with a
Coronal Mass Ejection (CME) which can cause aurora on planets
including the Earth \citep[see][for a review]{Fletcher2011}. 
  Indeed, it has been demonstrated that the most energetic (X class)
  X-ray flares are associated with a CME event
  (e.g. \citet{Yashiro2005}). Although detecting CME's from stars
other than the Sun are obviously much more difficult to identify,
there are some examples including one from Proxima Centauri
\citep{Zic2020}. However, some theoretical studies suggest that if a
star has a strong magnetic field it could suppress CMEs
(e.g. \citet{AlvaradoGomez2018}). However, given the high energy
  of the flare seen in MQ Dra and the solar flare-CME analogy, we
  consider it more likely that a CME was associated with this stellar
  flare.

We now determine how much material would be released {\it if} a CME
was associated with the flare we detect from the secondary star in MQ
Dra and compare that with the mass accretion rate onto the white
dwarf.  Using data largely derived from solar X-ray flares,
\citet{Aarnio2012} and \citet{Drake2013} determined the relationship
between flare energy and mass released from a CME (see also
\citet{Odert2017}). For a flare with energy $E_{bol}=2\times10^{35}$
erg (and converting to equivalent X-ray energy using an assumption
$E_{X}=E_{bol}/100$, see \citet{OstenWolk2015}) we find a mass of
$\sim2-4\times10^{18}$ g would be released. The mass accretion rate of
MQ Dra has been estimated to be $\sim$$6\times10^{-14}$ {${M_{\odot}}$
  yr$^{-1}$} \citep{Schmidt2005} which is equates to
$\sim9\times10^{15}$ g over 40 min. This is more than two orders of
magnitude smaller than the amount material expected to be released
from the CME.  This suggests that either a CME was not associated with
the flare (an option which we consider less likely), or that it was
directed in a direction which precluded it being accreted by the white
dwarf, or that if it was accreted by the white dwarf the radiation was
emitted predominantly at UV or X-ray wavelengths.

\section{Discussion}

Distinguishing short lived mass transfer events, which occur through
Roche Lobe overflow, from stellar activity from the secondary star is
difficult as high-cadence photometry is required to resolve the
  profile of the event. The majority of stellar flares show a sharp
  rise to maximum followed by a slower decline with the whole event
  typically lasting a few tens of min. A short duration accretion
  event is likely to have a more top-hat shaped profile.
Observations of a burst from UZ For in a low state using X-ray and UV
data from XMM-Newton, indicated it was likely due from an accretion
event rather than a coronal flare \citep{StillMukai2001}. In the
optical, high amplitude flares from Polars are very rare. Although
\citet{Shakhovskoy1993} reported a 2 mag flare from AM Her which
lasted 20 min, not a single event with an amplitude greater than 2 mag
was detected from AM Her in 13 years of RoboScope observations
\citep{Kafka2005}. (There was evidence for lower amplitude, 0.2-0.6
mag, flare-like events from AM Her in a low state). One other example is 
V358 Aqr which showed a
$\sim$6 mag flare in an intermediate state and had a profile entirely
consistent with a stellar flare \citep{Beuermann2017}.  What remained
uncertain was whether the stellar flare caused an accretion event onto
the white dwarf.

MQ Dra showed one 10$^{35}$ erg flare in 89.0 d of {\tess}
observations. How does this rate compare with other late type M dwarf
stars?  M dwarfs have been observed using {\sl Kepler}, {\sl K2} and
{\tess} to determine their rotation period and investigate flare
activity levels
\citep[e.g.,][]{Paudel2018,Doyle2019,Paudel2019,Gunther2020}.  MQ Dra
shows 10$^{35}$ erg class flares at a similar rate to 2M0837+2050
(M8V) and 2M0831+2244 (M9V) \citep{Paudel2019}. 
  \citet{Gunther2020} made a survey of $\sim$25k low mass stars using
  data from the first two sectors of the {\tess} survey and found that
  very few stars in the M4-M10V range show 10$^{35}$ erg class
  flares. In contrast the fraction of M dwarfs which show optical
  flares at all increases from M1V ($\sim$5 percent) to M6V
  \citep[$\sim$45 percent,][]{Gunther2020} implying high levels of
  flare activity at lower energy levels. This is likely related to the
  lifetime of a stars activity: using H$\alpha$ as a proxy for
  activity \citep{West2008} found the active lifetime of a M1V star
  was only a few 100 Myr but rises to $\sim$7 Gyr by M5V spectral
  type.

One of the principal factors in determining the degree of flare
activity is a stars age, with activity declining as stars get older,
e.g., \citet{Skumanich86} and more recently by \citet{Davenport2019}.
Indeed, for M dwarfs stars with rotation periods shorter than 10
  days, $\sim$60 percent show flares \citep{Gunther2020}. However,
  \citet{Gunther2020} also noted that for stars with rotation periods
  shorter than 0.3 d, the flare rate dropped. Using a larger sample
  from all sectors in the southern hemisphere, \citet{Ramsay2020}
  confirmed this, finding that only a tenth of stars with rotation
  periods shorter than 0.2 d showed flare activity. This is the
  rotation period we expect for secondary stars in CVs, which we now
  return to.

Stars in a binary system are expected to rotate very close to the
  binary orbital period, implying a low differential rotation, which
  is expected to play a key role in generating a magnetic field in
  solar type stars \citep{Scharlemann1982}. Observations of the
pre-CV V471 Tau, which has a late K spectral type secondary
\citet{Hussian2006}, was found to rotate as a solid body. Despite the
lack of differential rotation, the secondary star in V471 Tau is
observed to be active (e.g. \citep[e.g.,][and references
  therein]{Hakala1989}. However, observations of AE Aqr, which has a
rapidly rotating white dwarf and a M4V secondary star, showed it was
not fully locked and has small differential rotation
\citep{Hill2014}. More recent work has shown that a strong
  magnetic field can be generated in fully convective stars in a
  different way to solar type stars
  \citep[e.g.,][]{Yadav2015,Yadav2016,Klein2021} indicating that fully
  convective stars could have strong magnetic fields (implying
  activity) even if their surface differential rotation is low.

The secondary stars in Polars are fast rotators with nearly all having
rotation periods shorter than 0.2 d. The vast majority also fill their
Roche Lobe, with only a small number, such as MQ Dra, underfilling
their Roche Lobe.  We now know that high energy bursts have been
detected from Polars which fill and underfill their Roche
Lobe. However, the impact of a star filling its Roche Lobe upon its
differential rotation rate remains unclear. It remains to be seen what
fraction of low mass stars with rotation periods $<$0.2 d are in
binary systems. Could they show high energy flares if they were
observed over a sufficiently long time interval?

To better understand the activity levels in secondary stars in compact
binaries, we encourage high cadence photometric observations of Polars
in intermediate or low states to search for low amplitude as well as
the rare high energy events. In particular it is hoped that MQ Dra
will be observed again in {\sl TESS} Cycle 4 in either 2 min or 20 sec
mode.

\section*{Acknowledgements}

This paper includes data collected by the {\tess} mission. Funding for
the {\tess} mission is provided by the NASA Explorer Program.  The
Gaia archive website is
\url{https://archives.esac.esa.int/gaia}. Armagh Observatory and
Planetarium is core funded by the Northern Ireland Executive through
the Dept. for Communities. We thank the referee for a 
constructive and helpful report.

Data Availability: The {\tess}  data are available from
the NASA MAST portal.

\vspace{4mm}

\bibliographystyle{mnras}

\bsp	
\label{lastpage}

\end{document}